\documentclass[aps,pra,reprint,superscriptaddress,amsmath,showpacs,amssymb]{revtex4-1}
\usepackage{bm}
\usepackage[dvips,final]{graphicx}
\usepackage{amsmath,amssymb}

\newif\ifHIDEHIGHLIGNT
\HIDEHIGHLIGNTtrue 
\ifHIDEHIGHLIGNT
	\usepackage{ulem,color}
	\newcounter{nnn}

\else

\fi

\newcommand{\be}{\begin{equation}}
\newcommand{\ee}{\end{equation}}
\newcommand{\BM}{\begin{pmatrix}}
\newcommand{\EM}{\end{pmatrix}}

\renewcommand{\d}{\dagger}
\renewcommand{\phi}{\varphi}

\newcommand{\bra}[1]{\bigl\langle #1 \bigr|}
\newcommand{\ket}[1]{\bigl| #1 \bigr\rangle}

\newcommand{\intx}{\int\!d^3x\;}

\newcommand{\avg}[1]{\bra0 #1 \ket0}

\bmdefine{\bx}{x}
\bmdefine{\by}{y}
\bmdefine{\bz}{z}

\begin{document}
\title{Interacting multiple zero mode formulation and its application 
to a system consisting of a dark soliton in a condensate}

\author{J.~Takahashi}
\email{j.takahashi@aoni.waseda.jp}
\affiliation{Department of Electronic and Physical Systems, Waseda
University, Tokyo 169-8555, Japan} 
\author{Y.~Nakamura}
\email{yusuke.n@asagi.waseda.jp}
\affiliation{Department of Electronic and Physical Systems, Waseda
University, Tokyo 169-8555, Japan} 

\author{Y.~Yamanaka}
\email{yamanaka@waseda.jp}
\affiliation{Department of Electronic and Physical Systems, Waseda
University, Tokyo 169-8555, Japan} 

\date{\today}

\begin{abstract} 
To formulate the zero modes in a finite-size system 
with spontaneous breakdown of symmetries in quantum field theory is not trivial,
for in the naive Bogoliubov theory, one encounters difficulties such as phase diffusion, 
the absence of a definite criterion for determining the ground state, and infrared divergences.
A new interacting zero mode formulation that has been proposed for systems with 
a single zero mode to avoid these difficulties is extended to general systems with multiple
zero modes. It naturally and definitely gives the interactions among the quantized zero modes,
the consequences of which can be observed experimentally.
In this paper, as a typical example, we consider an atomic Bose--Einstein condensed system with
a dark soliton that contains two zero modes corresponding to spontaneous breakdown of the 
U(1) gauge and translational symmetries. Then we evaluate the standard deviations of the zero mode
operators and see how the mutual interaction between the two zero modes affects them.
\end{abstract}


\pacs{03.75.Hh, 03.75.Nt, 67.85.-d}

\maketitle

\section{Introduction}

Since the pioneering experiments~\cite{Ketterle,Cornell,Bradley},
the Bose--Einstein condensate (BEC) phenomenon in a trapped ultracold atomic system
has been the central subject of many experimental and theoretical studies.
We view the subject from the standpoint of quantum field theory, which is the most
fundamental dynamical law and in which the BEC in an atomic system is 
interpreted as a spontaneous breakdown of the U(1) gauge symmetry.

The concept of spontaneous symmetry breaking (SSB) 
yields many examples of successful descriptions of nature using quantum field theory.
In an infinite system with a spontaneously broken symmetry, the celebrated 
Nambu--Goldstone theorem~\cite{NGtheorem1,NGtheorem2}
implies the existence of a zero mode, reflecting the original symmetry.
The zero mode plays a crucial role in creating and retaining the ordered state.

Study of the BEC in a trapped system highlights the importance of 
treating the operators belonging to the zero mode sector carefully
because the system has a finite size, and the zero energy state also 
stands alone as a discrete level. The importance is easily overlooked
in a homogeneous infinite system, for the zero mode sector is buried in 
the continuum labeled by the momentum index. The practice of formulation, 
 usually called the Bogoliubov theory,
is to take an unperturbed Hamiltonian up to the
second power of a field operator  and to attempt to 
diagonalize it by expanding the field operator in the appropriate 
complete set. However, when the complete set includes eigenfunctions belonging 
to a zero eigenvalue, {\it i.e.}, when the system has zero modes, 
the unperturbed Hamiltonian of the zero mode sector cannot be diagonalized 
in terms of creation and annihilation operators but has the form of free particles
expressed by the quantum mechanical operators ${\hat P}$ and ${\hat Q}$.
We refer to this as the \textit{free zero mode formulation}. Although this part of the Hamiltonian
 is simply neglected for a homogeneous infinite system, it cannot be neglected 
for a finite-size system. Then, according to Ref.~\cite{Lewenstein}, 
the phase of the order parameter is definite only for a short time
 because ${\hat Q}$ is interpreted as a phase operator, and its
quantum fluctuation is given by 
$\Delta Q \sim t$ for large $t$. We also point out that there is no criterion for 
specifying a vacuum uniquely, as any energy eigenstate 
of a free particle has infinite $\Delta Q$. Thus,
it is concluded that the free zero mode 
formulation for a finite-size system is inconsistent.

To address this inconsistency, we proposed a new formulation for
spontaneous breakdown of the U(1) gauge symmetry in Ref.~\cite{ZeroState}.
The key point there
 is the inclusion of higher-than-third powers of ${\hat P}$ and ${\hat Q}$ 
in the unperturbed Hamiltonian, which yields their nonlinear equations of motion. 
We therefore call it the \textit{interacting zero mode formulation}. 
The stationary Schr\"odinger-like equation with the nonlinear 
unperturbed Hamiltonian of
 ${\hat P}$ and ${\hat Q}$ gives bound states rather than 
the free one, and the energy spectrum becomes
 discrete, so the ground state
is identified uniquely as the vacuum. Then $\Delta Q$ is independent of time,
and we have no inconsistency as long as the calculated $\Delta Q$ is small.

The interacting zero mode formulation not only enables us to 
describe quantum fluctuation of a zero mode properly, but also,
 when two or more symmetries are broken spontaneously and there 
are multiple zero modes, introduces interactions among the
 zero modes naturally.
In this paper, we focus on interactions among zero modes, extending
 the interacting zero mode formulation for a single zero mode \cite{ZeroState} 
to that for multiple ones. After giving a general formulation, 
we consider, as an example of its application, a system consisting of a dark soliton in a homogeneous
condensate, where two zero modes coexist 
corresponding to the spontaneously broken translational and U(1) gauge symmetries.
This example was studied in Ref.~\cite{Dziarmaga}, in which a 
nonperturbative treatment of the zero modes using an effective Hamiltonian 
was proposed. However, the resultant Hamiltonian represents the free motion of each
zero mode with no interaction, so our approach is essentially different
from that in Ref.~\cite{Dziarmaga}.

This paper is organized as follows:
In Sect.~II, we extend the interacting zero mode formulation for a single zero mode,
presented in \cite{ZeroState}, to a general case of multiple zero modes,
comparing it with the corresponding free zero mode case.
In Sect.~III, the general formulation is applied to the originally homogeneous
system containing a condensate and a dark soliton. Two zero modes appear and
interact with each other. We are particularly interested in the quantum fluctuations 
$\Delta Q_i$ (where $i$ labels each zero mode), which are affected by 
 interactions among the zero modes.
Section IV presents a summary and conclusion.
\bigskip

\section{FORMULATION OF MULTIPLE ZERO MODES IN QUANTUM FIELD THEORY}

In Ref.~\cite{ZeroState}, we proposed the interacting zero mode formulation 
for a finite-size system with spontaneous breakdown of the U(1) gauge symmetry.
The main motivation for the
 formulation is that the quantum fluctuation in the phase
of the order parameter cannot remain small in the conventional
free zero mode formulation, whereas its smallness is the starting 
assumption of the formulation. 
In this section, we extend the new formulation to cases of multiple zero modes.

We suppose a Hamiltonian with global U(1) gauge symmetry,
\be \label{eq:originalH}
	\hat{H} \!=\! \intx \! \left[ \hat{\psi}^\d \! \left(-\frac{\nabla^2}{2m} + V_{\mathrm{ex}} - \mu\right) \! \hat{\psi}
  + \frac{g}{2} \hat{\psi}^\d\hat{\psi}^\d\hat{\psi}\hat{\psi} \right] \,,
\ee
where $V_{\mathrm{ex}}$, $m$, $\mu$, and $g$ are the external potential, atomic mass,
chemical potential, and repulsive coupling constant ($g>0$), respectively. 
Throughout this paper, we set $\hbar$ to unity.
The bosonic field operator $\hat{\psi}$ obeys the canonical commutation relations 
$
	\bigl[ \hat{\psi}(\bx,t) , \hat{\psi}^\d(\bx',t) \bigr] = \delta(\bx-\bx') ,\,\,
	\bigl[ \hat{\psi}(\bx,t) , \hat{\psi}(\bx',t) \bigr] = 0 \,.
$
When the U(1) gauge symmetry is broken spontaneously, $\hat{\psi}$
is divided into coherent and incoherent parts as
 $\hat{\psi}=\xi + \hat{\phi}$, according to the criterion $\bra0 \hat{\phi} \ket0 = 0$.
The coherent part, or the order parameter $\xi$, is related to the total number 
of condensates, $N_0 = \intx |\xi|^2$, and the vacuum $\ket0$ is determined self-consistently later.
The total Hamiltonian is rewritten in terms of $\hat{\phi}$ as
$
	\hat{H} = \hat{H}_1 + \hat{H}_2 + \hat{H}_3 + \hat{H}_4 ,
$
where
\begin{align}
	\hat{H}_1 \!&=\! \intx \! \left[ \hat{\phi}^\d \! \left(-\frac{\nabla^2}{2m}+V_{\mathrm{ex}} - \mu + g|\xi|^2\right) \! \xi \right] \!+\!{\rm h.c.} \,,\\
	\hat{H}_2 \!&=\! \intx \! \left[ \hat{\phi}^\d \mathcal{L} \hat{\phi} + \frac12\hat{\phi}\mathcal{M}\hat{\phi} + \frac12\hat{\phi}^\d \mathcal{M}^* \hat{\phi}^\d \right] \,,\label{eq:defH2}\\
	\hat{H}_3 \!&=\! g\intx \xi \hat{\phi}^\d\hat{\phi}^\d\hat{\phi} + {\rm h.c.} \,,\\
	\hat{H}_4 \!&=\! \frac{g}{2}\intx \hat{\phi}^\d\hat{\phi}^\d\hat{\phi}\hat{\phi}\,,
\end{align}
where 
$\mathcal{L} = -\nabla^2/2m+V_{\mathrm{ex}} 
-\mu + 2g|\xi|^2\,$ and
$\mathcal{M} = g\xi^2\,$.

\subsection{FREE ZERO MODE FORMULATION}
In the conventional approach, one chooses $\hat{H}_1 + \hat{H}_2$ as the unperturbed Hamiltonian assuming small $\hat{\varphi}$ :
\be \label{def:HLW}
\hat{H}_0 = \hat{H}_1 + \hat{H}_2.
\ee
Because the vacuum of this Hamiltonian is time-independent, the field division criterion $\bra0 \hat{\phi}(\bx,t) \ket0=0$ and the Heisenberg equation yield $\hat{H}_1=0$.
This implies that $\xi$ should satisfy the Gross--Pitaevskii (GP) equation \cite{GP},
\be \label{eq:GP}
	\left(-\frac{\nabla^2}{2m}+V_{\mathrm{ex}} - \mu + g|\xi|^2\right)\xi=0\,.
\ee 
To diagonalize $\hat{H}_2$~\cite{Lewenstein,Matsumoto2}, we introduce the Bogoliubov--de Gennes (BdG) equation~\cite{Bogoliubov,deGennes} 
$
	T \by_n = \omega_n \by_n \,
$
with the doublet notations  
\be
	T = \BM \mathcal{L} & \mathcal{M} \\ -\mathcal{M}^* & -\mathcal{L} \EM  \,,\qquad 
	\by_n = \BM u_n \\ v_n  \EM\,.
\ee
Now, we restrict ourselves to cases where all the eigenvalues are real, 
which implies that the system is dynamically stable, in order to 
diagonalize the excited modes. The diagonalization of a system with complex modes 
(that is, a dynamically unstable system)
 is discussed in Ref. \cite{Mine}. We consider that some symmetries 
in addition to the U(1) gauge symmetry
are spontaneously broken and we need two or more
 eigenfunctions belonging to zero eigenvalue, {\it i.e.}, $T \by_{0,i}=0\,$ where $\by_{0,i} = (f_i ,\; -f_i^*)^t\,$, with a label $i$. For the sake of completeness,  one has to 
introduce an adjoint function $\by_{-1,i} = (h_i ,\; h_i^*)^t$ to each $\by_{0,i}$,
which satisfies 
$T\by_{0,i}=I_{i}\by_{-1,i}$ and 2$\intx \! h_i^* \,f_i=1\,,$ with normalization constants $I_i$.
Let us expand $\hat{\phi}(\bx,t)$ by the BdG complete set as $\hat{\phi}(\bx,t)= \hat{\phi}_0(\bx,t) + \hat{\phi}_{\text{ex}}(\bx,t)$, where
\begin{align} \label{eq:phi_expansion}
	\hat{\phi}_0(\bx,t)           &= \sum_{i=\text{z.m.}} \left[ -i\hat{Q}_i(t) f_i(\bx) + \hat{P}_i(t) h_i(\bx) \right] \,,\\
	\label{eq:phi_expansion2}
	\hat{\phi}_{\text{ex}}(\bx,t) &= \sum_{\ell=\text{ex.}} \left[ \hat{a}_\ell(t) u_\ell(\bx) + \hat{a}_\ell^\d(t) v_\ell^*(\bx) \right] \,,
\end{align}
where ``z.m.'' and ``ex.'' represent summations over the zero and excitation modes, respectively. 
The operators satisfy 
$
	[\hat{Q}_i(t), \hat{P}_j(t)] = i\delta_{ij} ,\,
	[\hat{a}_\ell(t), \hat{a}_{\ell'}^\d(t)] = \delta_{\ell\ell'} \,,
$
and the vanishing ones otherwise, where $\hat{Q}_i(t)$ and $\hat{P}_i(t)$, 
also called the zero mode operators or quantum coordinates, are hermitian.
Substituting the expansions (\ref{eq:phi_expansion}) and (\ref{eq:phi_expansion2}) into
 Eq.~(\ref{eq:defH2}), we obtain
\be \label{eq:H2}
	\hat{H}_0 = \hat{H}_2 =\sum_{i=\text{z.m.}}\frac{\hat{P}_i^2}{2I^{-1}_i} + \sum_{\ell=\text{ex.}} \omega_\ell \hat{a}_\ell^\d \hat{a}_\ell \,.
\ee
The unperturbed Hamiltonian~(\ref{eq:H2}) is represented by the sum of 
the harmonic part of the excitation for each $\ell$  and the free part of 
the quantum coordinates for each $i$. Because of the latter contribution, we call this conventional 
formulation the free zero mode one. As a matter of course, there is no interaction 
among the zero modes in the Hamiltonian~(\ref{eq:H2}).
The vacuum of the total system would be the ground state of Eq.~(\ref{eq:H2}): 
$
    \ket0 = \prod_{i}\ket0_i \otimes \ket0_{\text{ex}} \,, 
$
where $\ket0_i$ is the ground state in the $i$th zero mode sector 
 satisfying the field division criterion 
$
 { }_i\bra0 \hat{Q}_i \ket0_i =\, { }_i\bra0\hat{P}_i \ket0_i = 0 \,.
$

However, the choice of $\ket0_i$ causes problems.
First, $I_i$ may be negative, which corresponds to having
 a system with a negative ``mass,''  
and there is no ground state. On the other hand, for $I_i>0$, 
the lowest eigenstate is the zero ``momentum'' state, $\hat{P}_i\ket{0}_i=0$. 
As a result of the uncertainty relation, the standard deviation of $\hat Q_i$,
denoted by $\Delta Q_i=\sqrt{_i\bra0\hat{Q}^2_i \ket0_i - {}_i\bra0\hat{Q}_i\ket0_i^2}$,
diverges. This divergence immediately conflicts with 
the starting assumption
of small $\hat{\phi}$. It also yields an unphysical situation in which 
some physical quantities such as the total number density also diverge,
 which will be seen in Eq.~(\ref{meaningQx:depletion}). 
One way to resolve this contradiction might be to choose a wave packet state 
as the vacuum \cite{Lewenstein}. Although  $\Delta Q_i$ is finite at $t=0$, 
it is again divergent after a long time because of the collapse of the 
wave packet; that is,  $\hat{Q}_i(t)=\hat{Q}_i(0)+\hat{P}_i(0) t$ 
 and $\Delta Q_i \propto t$ for large $t$. It is argued in Ref.~\cite{Lewenstein} that
 $\hat{Q}$, acting as the phase operator of the 
order parameter, yields 
a divergent $ \Delta Q(t)$ as $t$ goes to infinity, and that the phase diffuses. All of the above
pathological properties are rooted in the fact that the free Hamiltonian
has a continuous spectrum.

We note that the difficulties concerning quantum fluctuations 
of zero modes are veiled in the Bogoliubov approximation,
in which the original creation and annihilation operators associated with
the eigenfunction belonging to zero eigenvalue
are replaced with classical numbers in the field ${\hat \psi}$,
or the zero mode operators in ${\hat \varphi}$ are simply neglected.

\subsection{INTERACTING ZERO MODE FORMULATION}

Although the choice of the unperturbed Hamiltonian~(\ref{def:HLW}) 
is based on the assumption of small $\hat{\phi}$, or, more precisely, 
small $\hat{\phi}_0$ and small $\hat{\phi}_{\text{ex}}$, 
$\avg{\hat{Q}_i^2(t)}$ is divergent, which indicates large $\hat{\phi}_0$.
Thus, the zero mode fluctuations cannot be kept small, and
the assumption of small $\hat{\phi}_0$ has to be abandoned. This 
 recognition is the starting
point of Ref.~\cite{ZeroState}. Gathering all the terms consisting only
of $\hat{\phi}_0$ in the total Hamiltonian, we introduce the 
new unperturbed Hamiltonian
\begin{align} \label{eq:Hu}
	\hat{H}_u = \hat{H}_2 + \Delta \hat{H} \,,
\end{align}
where $\hat{H}_{1}=0$, for the same reason as in the previous subsection.
The additional component, $\Delta \hat{H}$, is the sum of the third and fourth 
powers of the zero mode operators and their counter  terms,
\begin{align}
	\Delta \hat{H} = \hat{H}^{QP}_3 + \hat{H}^{QP}_4 \!-\! \sum_{i=\text{z.m.}} \left[ \delta\mu_i \hat{P}_i + \delta\nu_i \hat{Q}_i \right] \,.
\end{align}
The superscript $QP$ indicates that all the terms consisting only of
$\hat{Q}_i$ and $\hat{P}_i$ are picked up.
We set up the stationary Schr\"odinger-like equation,
\be \label{eq:HuEigen}
 \hat{H}_u^{QP} \ket{\Psi_\nu}= E_{\nu}\ket{\Psi_\nu} \, \quad
(\nu=0,1,2,\cdots) .
\ee
As was seen in Ref.~\cite{ZeroState} for the model with the single zero mode
and will be seen in the example with two zero modes in the next section, 
the eigenequation~(\ref{eq:HuEigen})
is a type of bound state problem and yields a discrete spectrum, in contrast to 
the free zero mode formulation in the previous subsection.
It is quite natural to take the whole unperturbed vacuum 
$\ket0= \ket{\Psi_0} \otimes \ket{0}_{\mathrm{ex}}$, where
$\ket{\Psi_0}$ is the ground state of the above equation.
The unknown parameters $\delta\mu_i$ and $\delta\nu_i$ involved in $\hat{H}_u^{QP}$
 should be determined 
so as to satisfy the field division criterion 
\be \label{eq:criterion_of_division}
	\bra{\Psi_0} \hat{Q}_i \ket{\Psi_0}=
	\bra{\Psi_0}\hat{P}_i \ket{\Psi_0} = 0 \,
\ee
in a manner consistent with Eq.~(\ref{eq:HuEigen}).

Substituting the expansion (\ref{eq:phi_expansion}) into Eq.~(\ref{eq:Hu}), we gather it as
$
	\hat{H}_u = \hat{H}_{u,1}^{QP} \!+\! \hat{H}_{u,2}^{QP} \!+\! \hat{H}_{u,3}^{QP} \!+\! \hat{H}_{u,4}^{QP} \!+\! \sum_\ell \omega_\ell \hat{a}_\ell^\d \hat{a}_\ell \,,
$
where
\begin{align}
 \hat{H}_{u,1}^{QP} &\!=\! -\delta\mu_{i} \hat{P}_{i} -\delta\nu_{i} \hat{Q}_{i} ,\, \\
 \hat{H}_{u,2}^{QP} &\!=\! \frac{\hat{P}^2_{i}}{2I^{-1}_{i}} ,\, \\
 \hat{H}_{u,3}^{QP} &\!=\! 2 \mathrm{Re} \Big[\!-\! i A_{\theta jk\ell} \hat{Q}_j \hat{Q}_k \hat{Q}_\ell
                                       \!+\!  B_{\theta jk\ell} \hat{Q}_j \{\hat{Q}_k,\, \! \hat{P}_\ell \} \nonumber\\
                                       & \hspace{1cm}
                                       \!-\!  B^*_{kj\theta\ell} \hat{P}_\ell \hat{Q}_k \hat{Q}_j
                                       \!+\! i C_{\theta jk\ell} \hat{Q}_j \hat{P}_k \hat{P}_\ell        \nonumber\\
                                       & \hspace{1cm}
                                       \!-\! i C'_{\theta jk\ell} \hat{P}_k \{\hat{Q}_j,\, \! \hat{P}_\ell \}
                                       \!+\! D_{\theta jk\ell} \hat{P}_j \hat{P}_k \hat{P}_\ell
                                       \Big] ,\,\label{eq:Hu3}\\
 \hat{H}_{u,4}^{QP} &\!=\! \frac{A_{ijk\ell}}{2} \hat{Q}_i \hat{Q}_j \hat{Q}_k \hat{Q}_\ell 
                         \!-\! \mathrm{Im} \! \left[ B_{ijk\ell} \right] \hat{Q}_i \hat{Q}_j \{\hat{Q}_k,\, \! \hat{P}_\ell \}  \nonumber\\
                         & \hspace{0.4cm} 
                         \!-\! \mathrm{Re} \! \left[ C_{ijk\ell} \right] \hat{Q}_i \hat{Q}_j \hat{P}_k \hat{P}_\ell  
                         \!+\! \frac{C'_{ijk\ell}}{2} \{ \hat{Q}_i,\, \! \hat{P}_k \} \{ \hat{Q}_j,\, \! \hat{P}_\ell \}          \nonumber\\
                         & \hspace{0.4cm} 
                         \!-\! \mathrm{Im} \! \left[ D_{ijk\ell} \right] \{ \hat{Q}_i,\, \! \hat{P}_j \} \hat{P}_k \hat{P}_\ell
                         \!+\! \frac{E_{ijk\ell}}{2} \hat{P}_i \hat{P}_j \hat{P}_k \hat{P}_\ell  \,,                        \label{eq:Hu4}
\end{align}
with
\begin{alignat}{3} \label{eq:defAE}
A_{ijk\ell}   &= g\int dx f^*_i f^*_j f_k f_\ell ,\,\, 
B_{ijk\ell}   = g\int dx f^*_i f^*_j f_k h_\ell ,\,\,   \nonumber\\
C_{ijk\ell}  &= g\int dx f^*_i f^*_j h_k h_\ell ,\,\,
C'_{ijk\ell}  = g\int dx f^*_i f_j h^*_k h_\ell ,\,\,   \nonumber\\
D_{ijk\ell}  &= g\int dx f^*_i h^*_j h_k h_\ell ,\,\,
E_{ijk\ell}    = g\int dx h^*_i h^*_j h_k h_\ell .     
\end{alignat}
In Eqs.~(\ref{eq:Hu3}) and (\ref{eq:Hu4}), we define 
$\{\hat{O}_1,\hat{O}_2\}\equiv\hat{O}_1\hat{O}_2 + \hat{O}_2\hat{O}_1$,
 and the dummy indices should be summed over.
A remarkable consequence of the present formulation is that 
we have cross-terms among the different zero mode operators in the unperturbed
Hamiltonian, namely, interactions and mixings among them.
In other words, the Hamiltonian ${\hat H}_u$ is an effective Hamiltonian 
in the zero mode sector governing the dynamics of the condensate and is uniquely
derived from the original Hamiltonian~(\ref{eq:originalH}).

\section{APPLICATION TO HOMOGENEOUS SYSTEM WITH DARK SOLITON}
\begin{figure}[tbh!]
\begin{center}
\vspace{1cm}
\includegraphics[width=8cm]{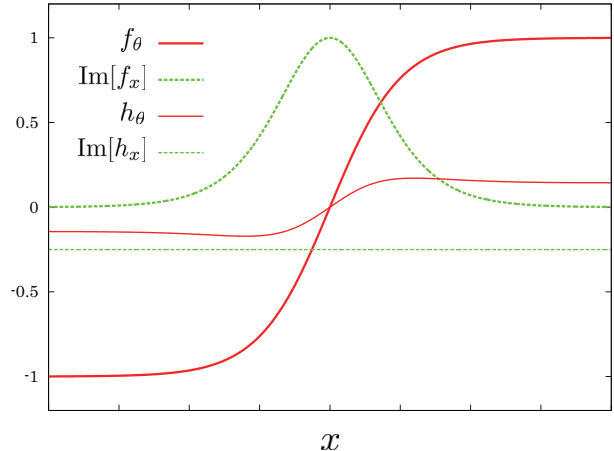}
\end{center}
\caption{\footnotesize{%
(Color online) Plots of the zero modes and their adjoint eigenfunctions.
Red bold and thin solid lines indicate $f_\theta$ and $h_\theta$, respectively.
Green bold and thin broken lines indicate $f_x$ and $h_x$, respectively.
}}
\label{figfh}
\end{figure}

In this section, we consider a system consisting of a dark soliton 
in a homogeneous system, which is described by the Hamiltonian
in Eq.~(\ref{eq:originalH}) with $V_{\mathrm{ex}}=0$ and has two zero modes.
The GP solution of a one-dimensional dark soliton is
\begin{align}
\xi(x) = \sqrt{n_0} \tanh \left\{\kappa (x-x_0) \right\},\quad \mu = gn_0 \, ,
\end{align}
where $\kappa = \sqrt{mg n_0}$, and $n_0$ is the bulk density of the condensate.
Hereafter, we set $x_0=0$ and $n_0=1$ for the sake of simplicity.
The zero modes and their adjoint eigenfunctions are
\begin{align}
f_{\theta} &= \xi(x) \,, \label{def_of_f_t} \\
f_{x}      &= i\frac{d}{dx} \xi(x) \,, \label{def_of_f_x}\\
h_{\theta} &= \frac{\sqrt{g}}{2L} \left[ \tanh (\kappa x) \!+\! \kappa x \left\{ 1 \!-\! \tanh^2 (\kappa x) \right\} \right] \,,\\
h_{x}      &= -\frac{i}{4} \,,\label{def_of_h_x}\\
I_\theta   &= \frac{g}{L} \,, \,\,\,\, I_x = -\frac{g}{4\kappa} \,,
\end{align}
where the subscripts $\theta$ and $x$ denote the
$U(1)$ gauge and translational modes, respectively.
We plot the four functions for later convenience in Fig.~\ref{figfh}.
Note that although in Ref.~\cite{Dziarmaga},
which considered a nonperturbative treatment of the two zero modes, 
the system is put in an artificial box with a size $L$ and the boundary condition 
$f_\theta(-L/2)=f_\theta(L/2)=0$, our boundary condition is antiperiodic, 
$f_\theta(-L/2)=-f_\theta(L/2)\neq 0 $.
The operators $\hat{Q}_\theta$ and $\hat{Q}_x$, which are associated with the eigenfunctions
 (\ref{def_of_f_t}) and (\ref{def_of_f_x}), respectively, may
 be interpreted as the phase and  position operators
 of the soliton,
\begin{align} \label{meaningQx:position}
\hat{\psi}(x) &= \xi(x) - i\hat Q_\theta \xi(x) + \hat{Q}_x \frac{d\xi(x)}{dx} + \cdots \nonumber\\
              &\simeq \xi(x+\hat{Q}_x)e^{-i\hat{Q}_\theta}.
\end{align}
However, this interpretation is true only 
when $\hat{Q}_\theta$ and $\hat{Q}_x$ are small. Our approach does not rely on it.

\subsection{FREE ZERO MODE APPROACH FOR DARK SOLITON}
In the conventional treatment, the unperturbed Hamiltonian (\ref{eq:H2}) for this system is described by
\be \label{def:HDZ}
	\hat{H}_0 = \frac{\hat{P}_\theta^2}{2I^{-1}_\theta} + \frac{\hat{P}_x^2}{2I^{-1}_x} +\sum_{\ell=\text{ex.}} \omega_\ell \hat{a}_\ell^\d \hat{a}_\ell \,,
\ee
with positive $I_\theta$ and negative $I_x$.
As mentioned above, the ground state for the
U(1) gauge zero mode induces phase diffusion, and there is no ground state 
for the translational zero mode because of the negative ``mass,'' $I_x<0$.
In Ref.~\cite{Dziarmaga}, Eq.~(\ref{def:HDZ}) is derived from the classical 
Lagrangian for the collective coordinates as the starting point of the 
nonperturbative treatment, and  the U(1) gauge zero mode sector is 
restricted to a subspace with a definite total number of atoms, assuming 
a large phase fluctuation. 
After this procedure, the calculated $\Delta Q_x$ with a Gaussian wave packet state
grows with $t$.
As can be seen from the total number density 
\begin{align} \label{meaningQx:depletion}
\bra0 &\hat{\psi}^\d(x,t) \hat{\psi}(x,t) \ket0 \nonumber\\ 
  &= \left[ 1+ \Delta Q_\theta^2(t) \right] |\xi(x)|^2 \!+\! \Delta Q_x^2(t) \Big| \frac{d\xi(x)}{dx} \Big|^2 + \cdots \,,
\end{align}
$\Delta Q_x$ is related to the number of atoms
that fill the center of the soliton. 
It has been indicated that $\Delta Q_x(t)$ grows 
with $t$, which is quantum depletion.
A similar result of quantum depletion 
for a system with an attractive interaction having
a bright soliton was obtained in Ref.~\cite{Huang}, 
although then the ``mass'' $I_x$ is positive.

The difficulty of the free zero mode approach lies in the 
the negative ``mass'' problem, the absence of a ground state for the unperturbed Hamiltonian~(\ref{def:HDZ}), and the fact that $\Delta Q_x(t)$ increases to infinity with time.

\subsection{INTERACTING ZERO MODE APPROACH FOR DARK SOLITON}
In the interacting zero mode approach, we 
can obtain the natural vacuum, which 
causes neither phase dissipation nor
 the negative ``mass'' problem, by solving the Schr\"{o}dinger-like 
equation~(\ref{eq:HuEigen}) with the unperturbed Hamiltonian~(\ref{eq:Hu}).

\begin{figure}[tbh!]
\begin{center}
\vspace{1cm}
\includegraphics[width=8cm]{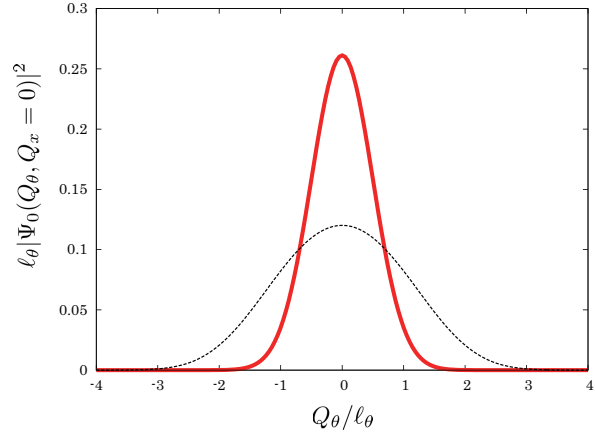}
\end{center}
\caption{\footnotesize{%
(Color online) Ground state distribution 
$|\Psi_0(Q_\theta, Q_x=0)|^2$ with and without the mutual interaction for $g=1$ and $L=1000$.
 (Red) bold solid and thin  broken lines indicate the distributions
for $\hat{H}^{QP}_u$ and $\hat{H}^{QP}_{u,\theta} + \hat{H}^{QP}_{u,x}$, respectively.
 Axes are scaled by $\ell_\theta = \,^6 \! \sqrt{I_\theta/24 A_{\theta\theta\theta\theta}}$ (see Ref.~\cite{ZeroState}).
}}
\label{Fig:groundstate}
\end{figure}

First, one has the nonperturbative Hamiltonian 
consisting only of ${\hat Q}_i$
and ${\hat P}_i$ with $i$ equal to either $\theta$ or $x$, denoted by $\hat{H}^{QP}_{u,i}$
$(i=\theta\,,\,x)$. In addition, there are interactions between 
$\{{\hat Q}_\theta\,,\,{\hat P}_\theta\}$ and $\{{\hat Q}_x\,,\,{\hat P}_x\}$,
on which we focus our attention. We seek the dominant contribution of the interaction
terms in the present model.
 Equations~(\ref{def_of_f_t})--(\ref{def_of_h_x}) and Fig.~\ref{figfh} show that $f_\theta$ and
$h_\theta$ are odd functions with respect to the variable $x$, whereas $f_x$ and
$h_x$ are even ones. The indices $(i,j,k,\ell)$ of
the non-vanishing cross-term coefficients in Eq.~(\ref{eq:defAE}) must be $(\theta,\theta,x,x)$
in random order. We consider the limit of large $L$,
much larger than the coherent length.
Then the magnitude of $h_x$ is small, {\it i.e.}, $1/L$, so
the $D$ and $E$ terms can be neglected.
The function $f_x$ peaks sharply around $x=0$, 
and the contributions of the $A$ and $B$ terms are small. 
Thus, the dominant contributions come only from the $C$ and $C'$ terms.
One can neglect the third-power terms with respect to the
 zero mode operators because their contributions are small compared 
with those of the fourth-power terms owing to the field division criterion. 
Consequently, we have the approximate Hamiltonian
\begin{align} \label{Eq:Heff}
\hat{H}^{QP}_u
&\simeq\hat{H}^{QP}_{u,\theta} + \hat{H}^{QP}_{u,x} \!+\! 3|C_{\theta\theta xx}| \hat{Q}^2_{\theta}\hat{P}^2_{x} \,.
\end{align}
Here the last term represents the dominant interaction between the U(1) 
gauge and translational zero modes.
Solving the Schr\"{o}dinger-like equation 
(\ref{eq:HuEigen}) with the Hamiltonian 
(\ref{Eq:Heff}) numerically, 
we find the ground state or vacuum in the zero mode sector.
To illustrate the significance of the mutual interaction, namely, the last term in
Eq.~(\ref{Eq:Heff}), we plot the ground state distribution $|\Psi_0(Q_\theta, Q_x=0)|^2$
with and without it in Fig.~\ref{Fig:groundstate}; the difference between the plots
is striking. The sharp distribution in the presence of the mutual interaction can be understood
from the fact that the last term, behaving as $Q_\theta^2$,
 serves as an additional harmonic potential for $Q_\theta$.

\begin{figure}[tbh!]
\begin{center}
\vspace{1cm}
\includegraphics[width=8cm]{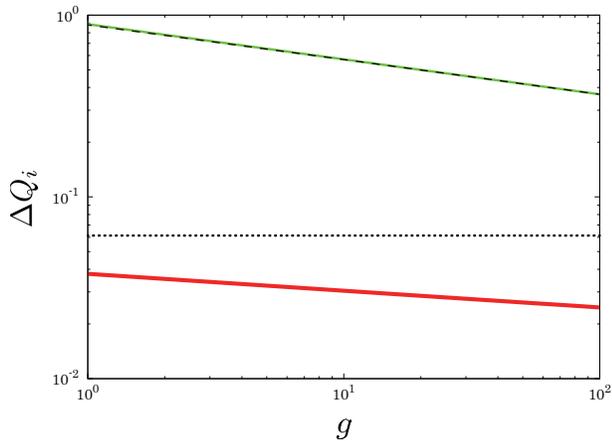}
\end{center}
\caption{\footnotesize{%
(Color online) $g$ dependences of standard deviations $\Delta Q_\theta$ 
and $\Delta Q_x$ with $m=1$ and $N_0=10^3$.
(Red) bold and (green) thin solid lines indicate $\Delta Q_\theta$ and $\Delta Q_x$ 
 for the ground state of $\hat{H}^{QP}_u$ in Eq. (\ref{Eq:Heff}), respectively.
Broken and dotted lines indicate the same quantities 
for the ground state of the Hamiltonian
$\hat{H}^{QP}_{u,\theta} + \hat{H}^{QP}_{u,x}$, namely, for the ground state without the 
mutual interaction between the two zero modes.
}}
\label{figSDg}
\end{figure}

One can evaluate the standard deviations $\Delta Q_i$ 
using the ground state distribution obtained above.
The results versus the parameter $g$ are presented in Fig.~\ref{figSDg},
which shows that $\Delta Q_\theta$ and $\Delta Q_x$ decrease 
according to the power law $g^{-\alpha_i}$, with $\alpha_\theta = 0.092\cdots$ and 
$\alpha_x=-0.192\cdots$.
It is natural that $\Delta Q_x$ decreases as $g$ increases, as
 the coherent length is proportional to $g^{-1/2}$ and the soliton becomes sharper for larger $g$.
We emphasize that the mutual interaction between the two zero modes 
qualitatively changes the $g$ dependence of $\Delta Q_\theta$: There
would be no $g$ dependence without the mutual interaction,
as in a normal condensate~\cite{ZeroState}.
The suppression of $\Delta Q_\theta$ because of the mutual interaction
may be rephrased as the condensate phase 
becoming rigid because of the presence of a
soliton with a $\pi$ phase kink.

\begin{figure}[tbh!]
\begin{center}
\vspace{1cm}
\includegraphics[width=8cm]{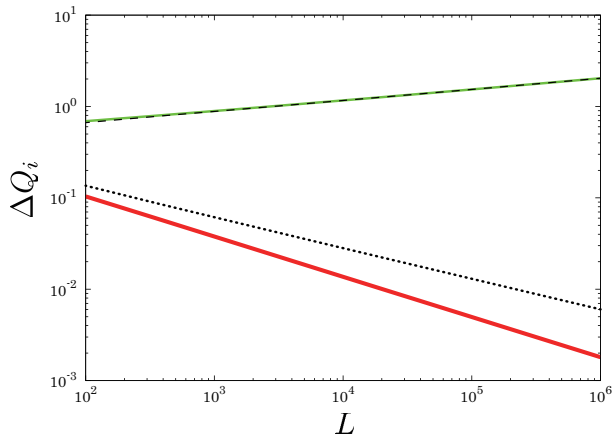}
\end{center}
\caption{\footnotesize{%
(Color online) Effect of system size on standard deviations $\Delta Q_\theta$ and $\Delta Q_x$ 
with $m=1$ and $g=1$.
Each line indicates the same quantity as in Fig.~\ref{figSDg}.
}}
\label{figSDL}
\end{figure}

So far we have considered a homogeneous limit, {\it i.e.}, the large $L$ limit.
We turn our attention to finite-$L$ systems and estimate the $L$ dependence,
for $L$ is a controllable parameter in experiments, although in that case 
the translational symmetry is broken not spontaneously but explicitly.
The calculations of the interacting zero mode formulation presented above 
are straightforwardly applied to a cylindrical system with a circumferential
 length $2L$ and two dark solitons, or to a system consisting of a dark soliton 
confined in a box of size $L$. For the latter, although
our antiperiodic boundary condition is not realistic, we expect that 
the calculations reflect the effects of finite $L$ qualitatively.
The $L$ dependence of 
$\Delta Q_i$ is shown
in Fig.~\ref{figSDL}, 
which shows that 
$\Delta Q_\theta$ decreases as $L^{\beta_\theta}$
 with $\beta_\theta= -0.440\cdots$, whereas $\Delta Q_x$ increases as $L^{\beta_x}$ with
$\beta_x=0.115\cdots$.
It is not surprising that $\Delta Q_\theta$ depends on $L$, 
as the eigenfunction $f_\theta(x)$ 
in Eq.~(\ref{def_of_f_t}) extends over the entire range 
from $-L/2$ to $L/2$. On the other hand, the fact 
that $\Delta Q_x$ also clearly depends on $L$ is puzzling at first glance,
because the eigenfunction $f_x(x)$ in Eq.~(\ref{def_of_f_x}), with which 
the position operator of the soliton $\hat Q_x$ is associated,
has a sharp distribution around the center of the soliton and 
is not affected by $L$.
The puzzle can be resolved as follows:
Whereas $\hat{Q}_x$ describes the position of the soliton, 
its conjugate partner $\hat{P}_x$ corresponds to its momentum or velocity
and is size-dependent, as the spread of $h_x(x)$ is of order $L$
[see Eqs.~(\ref{eq:phi_expansion}) and (\ref{def_of_h_x})].
 It is well known that although a standing soliton has 
a $\pi$ phase kink, a moving one has a smaller kink~\cite{Soliton_Review}. 
In view of this, the global character of $h_x(x)$, 
as seen in Eq.~(\ref{def_of_h_x}) to untwist the phase kink, could be understood. 
Because $\Hat{Q}_x$ and $\Hat{P}_x$ are canonically conjugate to each other, and one has
the uncertainty relation $\Delta Q_x \cdot \Delta P_x \sim 1/2$, 
the standard deviation 
$\Delta Q_x$ depends on $L$ as a consequence of the $L$ dependence
of $\Delta P_x$.
The size effects  may be observed in experiments on finite-size systems
in which $L$ is controlled.

\section{Summary}

Considering that the zero mode is the essence of SSB  and
that its quantum fluctuation must be treated properly,
we adopted the interacting zero mode formulation and 
extended it from a single zero mode system to a general case 
of multiple zero modes. It yields the effective Hamiltonian
of a pair of canonical operators for each zero mode,
the spectrum of which is discrete, as in the case of a single zero mode system, and 
 introduces interactions among the zero modes naturally and definitely.
The physical picture of zero modes interacting with each other is 
quite new. 

As an application of the new formulation,  
a system of size $L$ with a dark soliton is considered. 
In the large $L$ or homogeneous limit, there are two zero modes
corresponding to spontaneous breakdown of the 
U(1) gauge and translational symmetries. We investigated this system 
by performing calculations. The vacuum is obtained uniquely, and 
the standard deviations for the zero mode operators
$\Delta Q_i$ can be evaluated. The mutual interaction between the two zero
modes influences the ground state distribution and therefore $\Delta Q_\theta$,
and its effect is seen in the way that $\Delta Q_i$ depends on the coupling constant $g$.

As the trapped ultracold atomic system has a finite size $L$ 
in real experimental situations, we also studied the $L$ dependence of $\Delta Q_i$,
keeping $L$ finite in our calculations. The results may be checked in experiments
on a cylindrical system with two dark solitons or a dark soliton system confined
in a finite region.

As a characteristic of soliton physics, 
the behavior of the translational zero mode of the dark soliton
is expected to correlate with its velocity. A study of the correlation is a future work.

\begin{acknowledgments}
This work is supported in part by a Grant-in-Aid for Scientific Research (C) (No. 25400410) from the Japan Society for the 
Promotion of Science, Japan; ``Ambient SoC Global Program of Waseda University'' of the Ministry of Education, Culture, 
Sports, Science and Technology, Japan; and Waseda University Grant for Special Research Projects (Project No. 2013B-102). 
\end{acknowledgments}

\end{document}